# An unusual isotope effect in a high-transition-temperature superconductor

G.-H. Gweon[1], T. Sasagawa[2,3], S. Y. Zhou[4], J. Graf[1], H. Takagi[2,3,5], D.-H. Lee[1,4] & A. Lanzara[1,4]

[1]*Materials Sciences Division, Lawrence Berkeley National Laboratory, Berkeley, California 94720, USA*

[2]*Department of Advanced Materials Science, University of Tokyo, Kashiwa, Chiba 277-8561, Japan*

[3]*CREST, Japan Science and Technology Agency, Saitama 332-0012, Japan*

[4]*Department of Physics, University of California, Berkeley, California 94720, USA*

[5]*RIKEN (The Institute of Physical and Chemical Research), Wako 351-0198, Japan*

**In conventional superconductors, the electron pairing that allows superconductivity is caused by exchange of virtual phonons, which are quanta of lattice vibration. For high-transition-temperature (high-$T_c$) superconductors, it is far from clear that phonons are involved in the pairing at all. For example, the negligible change in $T_c$ of optimally doped $Bi_2Sr_2CaCu_2O_{8+d}$ (Bi2212; ref. 1) upon oxygen isotope substitution ($^{16}O \circledR ^{18}O$ leads to $T_c$ decreasing from 92 to 91 K) has often been taken to mean that phonons play an insignificant role in this material. Here we provide a detailed comparison of the electron dynamics of Bi2212 samples containing different oxygen isotopes, using angle-resolved photoemission spectroscopy. Our data show definite and strong isotope effects. Surprisingly, the effects mainly appear in broad high-energy humps, commonly referred to as 'incoherent peaks'. As a function of temperature and electron momentum, the magnitude of the isotope effect closely correlates with the superconducting gap— that is, the pair binding energy. We suggest that these results can be explained in a dynamic spin-Peierls picture[2], where the singlet pairing of electrons and the electron–lattice coupling mutually enhance each other.**

We compare angle-resolved photoemission spectroscopy (ARPES) data of optimally doped Bi2212 samples at the three different stages of the isotope substitution loop $^{16}O \rightarrow ^{18}O \rightarrow ^{16}O$ (Supplementary Methods). In this way, we study directly the impact on the electron spectral function due to a modification of phonon properties, and thus gain insights into the nature of electron–phonon interaction in this material. Here we use the term 'phonons' loosely to denote quanta of lattice vibrations including spatially localized ones[3,4]. To ensure that material properties unrelated to the isotope mass did not change during the isotope substitution process, we controlled the sample growth condition with high precision (Supplementary Methods) and checked the sample quality with various post-growth characterization tools, including ARPES itself (Supplementary Figure). All ARPES data were recorded at the Advanced Light Source as detailed elsewhere[5].

Figure 1 shows low temperature (25 K) ARPES spectra and their dispersions along the nodal (Γ−Y) direction, where the superconducting gap is zero. In this and the rest of the figures, blue, red and green represent results for $^{16}$O, $^{18}$O and re-substituted $^{16}$O$_R$ ($^{18}$O→$^{16}$O) samples, respectively. In Fig. 1a, raw ARPES spectra as a function of energy (that is, the energy distribution curves, EDCs) are shown for different momenta along the nodal direction. Each EDC shows a peak, which sharpens up as momentum ($k$) approaches the normal state Fermi surface[6]. We loosely refer to a sharp peak as a 'coherent peak' (CP), and a broad hump as an 'incoherent peak' (IP)[7]. We have no intention of implying that the energy and the width of CPs satisfy the Landau quasiparticle requirement. Making a comparison between the blue and red curves in Fig. 1a, we detect definite but small isotope effect. The most notable change is the shift of the peak position in curves 4 and 5 by approximately 15 meV, which is bigger than our error bar by a factor of three. Inspecting curves 1–6 in Fig. 1a, we note that the isotope effect is maximum for binding energy in the range of 100–300 meV, and vanishes as energy decreases (cut 1) and increases (cut 6) from this region. We note that this energy range coincides with range of $J$ to 2$J$, where $J$ is the super-exchange interaction of neighbouring spins.

Figure 1b shows the dispersions derived from the ARPES spectra at fixed energies as a function of momentum, known as the momentum distribution curves (MDCs). As reported earlier[6,8–12], we observe a 'kink', that is, a change of the slope in the dispersion, at energy ~70 meV. This feature has been used as evidence that a bosonic mode renormalizes the electron dynamics[13]. A comparison between the $^{16}$O and $^{18}$O dispersions in Fig. 1b clearly shows that the kink separates the low-energy regime where the spectra show CP and negligible isotope effect, from the high-energy regime where the spectra show IP and appreciable isotope effect. This observation suggests that phonons contribute appreciably to the electron self energy. To illustrate the size of our error bar, we show the dispersion of the $^{18}$O→$^{16}$O re-substituted sample as the green line in Fig. 1b. Clearly the experimental uncertainties caused by isotope substitution are smaller than the isotope-induced changes reported here. We note that the insensitivity to isotope substitution at low energies is consistent with the notion of 'universal nodal quasiparticle properties' inferred from the doping independence of nodal Fermi velocity[11]. The marked increase of the EDC width above the kink energy (Fig. 1a and ref. 6) and the concurrent strengthening of the isotope effect suggest that this broadening is at least partly due to scattering of electron by phonons. Figure 1b inset shows the real part of the electron self energy ($\Sigma$) as a function of energy ($\omega$), Re$S$($w$), obtained from measuring the deviation of the experimental dispersion from a common straight line representing the band structure dispersion[14]. We regard Re$S$($w$) mainly as a tool to zoom in on the isotope-induced changes in the dispersion. As expected, the overall isotope-induced change in Re$S$($w$) extends up to very high energy. Additionally, the maximum position of Re$S$($w$), that is, the kink energy[14], shows an isotope-induced redshift.

Both our EDC and MDC analyses show that phonons have important effect on the electron dynamics in the energy range 100–300 meV. We note that in the same energy range it is believed that magnetic fluctuations also contribute importantly to the electron self energy[15]. We also note that the fact that the isotope effect of Re$S$($w$) extends up to very high energy goes well beyond the Migdal–Eliashberg model, where the phonon effect is confined near the kink energy[16].

In Fig. 2a, we report the evolution of the isotope effect as the electron momentum approaches the antinodal region near the M point, where the *d*-wave superconducting gap reaches its maximum (see Fermi surface diagram in Fig. 2d inset). For cuts 2–6, where a non-zero superconducting gap exists, the MDC dispersion is shown only up to the gap edge. We defer the discussion of the isotope effect on the gap to later in this Letter. At low energy, a kink in the dispersion can be clearly identified for all cuts (see arrows), becoming stronger for near-antinodal cuts[5,12,17]. For all cuts, the kinks show an isotope induced redshift, (5–10)±5 meV. As in Fig. 1, the kink energy defines a crossover from the low-energy regime where the spectra show CP and negligible isotope effect, to the high-energy regime where the spectra show IP and strong isotope effect. Again, this result suggests that phonons are indeed key players in causing the kink[5,6,17]. At higher energy, the isotope-induced changes increase significantly as the momentum gets closer to the antinodal region. Moreover, a subtle sign change of the isotope effect is observed near cut 3. The results in Fig. 2a are confirmed by raw MDC data—for example, those shown in Fig. 2b. As noted earlier, isotope-induced changes are fully reversible upon isotope re-substitution (green line).

In Fig. 2c we show the raw EDC spectra at the momentum value marked with a grey-filled circle in Fig. 2d inset. Here a shift of about - 30 meV of the IP is observed. This value is a factor of 2 larger than the maximum shift reported along the nodal direction (see cut 1 and Fig. 1a). The weak low energy CP in each EDC of Fig. 2c is the well-known weak superstructure (SS) replica of the main band CP (also, see Supplementary Figure). Their isotope shift (about - 6 meV) is an order of magnitude smaller than that of the IP at higher energy. Clearly, this strongly energy-dependent isotope effect cannot be ascribed to an overall spectral shift or be described by the Migdal–Eliashberg theory. The increase of the spectral shift with increasing binding energy suggests that the multi-phonon effect[18,19] contributes to the electron self energy up to very high energy[20]. Again, we note that in the high-energy regime not only the electron–electron interaction[21,22] but also the electron–phonon interaction[23] affect the electron dynamics.

The inset of Fig. 2a summarizes the momentum dependence of high-energy isotope shift. The isotope-induced shift is plotted as a function of the superconducting gap. The linear correlation suggests a strongly anisotropic isotope effect, which increases as the electron momentum approaches the antinodal region.

In Fig. 2d, we show the dispersion for a cut parallel to the M–Y direction, determined from the EDC peak position. The location of this cut is shown as cut 7 in the inset. As in Fig. 2a, it is remarkable that, the higher the binding energy is, the stronger is the isotope effect. As the bottom of the dispersion is reached (yellow-filled circles), an isotope shift (about - 40 meV) nearly 20% of the entire width of the dispersion is observed (see arrows). The increase of the shift is consistent with Fig. 2a inset.

An additional important point is the isotope effect on the superconducting gap, $D_k$ (see, for example, cuts 3–6 in Fig. 2a). Unlike the sample-independent, reproducible and reversible isotope effect at energies much larger than the gap energy, the changes of the superconducting gap are small and random in both magnitude and sign. In particular, the maximum value of the gap varies by about ±5 meV from one sample to another, regardless of the isotope mass. This suggests that the main cause of gap modification is disorder[24], not the change in isotope mass. This is very different from the behaviour of conventional *s*-wave superconductors[25].

In Fig. 3, we study the effect of isotope substitution above the superconducting transition temperature ($T$=100 K). Figure 3a shows the MDC dispersions along cuts 1 and 3–6 of Fig. 2a. Comparing Fig. 3a with Fig. 2a, we note an overall decrease of the isotope-induced changes. For example, in the antinodal region (that is, cuts 5 and 6) the isotope effect is markedly different below and above $T_c$, consistent with the raw MDCs and EDCs in Fig. 3b, c. This interesting finding—that the strength of electron–lattice interaction is strongly temperature dependent—invalidates previous assumptions[9,10,15] that a strong temperature dependence of spectroscopic features rules out phonons as explanation. As the last detailed point, we note that the sign reversal of the high-energy isotope effect near cut 3 persists at high temperature, despite the reduced overall isotope effect.

Before discussing the implications of our results, we address two possible sources of experimental error: unintentional doping change induced by the substitution process, and sample misalignment. Both of them are ruled out by the temperature dependence and by the reversibility and the reproducibility of the observed effects upon repeated measurements. Two more arguments can be used to rule out doping differences (also, see Supplementary Figure). First, a direct comparison with the MDC dispersion at different doping shows that the sign reversal at high energy cannot be induced by a doping change. Second, if the changes in the high-energy nodal dispersions in Fig. 1 are due to a doping change $\Delta x$, $\Delta x \gg 0.05$ (ref. 11) is implied, which is five times bigger than the maximum doping uncertainty ($\Delta x \approx 0.01$) of the isotope substitution process.

A potentially simple explanation of our findings is the effect of static lattice distortion, induced by the isotope substitution. However, it is unlikely that the small change in the oxygen mass can affect the overall lattice structure, as shown by diffraction results for $La_{2-x}Sr_xCuO_4$ (ref. 26). Moreover, the fact that the isotope effects occur with opposite signs for different cuts in momentum space excludes a net average lattice change as explanation. Finally, the isotope substitution might induce a random static structural disorder, connected with the disorder seen in scanning tunnelling microscopy[24]. However, whereas the observed fluctuation of the superconducting gap is consistent with this possibility, the larger and fully reproducible isotope-induced changes at high energy are not.

On the basis of these findings we propose the following model for the nature of electron–lattice coupling in high-$T_c$ superconductors. The fact that the isotope effect on the ARPES spectral function becomes much stronger below $T_c$ suggests a picture where pairing of electrons enhances their coupling to the lattice and vice versa, as in spin-Peierls physics[2]. In this picture, the motion of electron pairs modifies the lattice distortion locally. If the coupling between the electron pair and the lattice is too strong, the pair will be localized and the lattice distortion becomes static. In that limit, the system becomes an insulator rather than a superconductor. In the interesting situation where the coupling is not that strong, the dynamic spin-Peierls distortion follows the coherent motion of electron pairs in a superconducting state. Creation of a photo-hole in this state causes the lattice to lose its distortion, and a strong coupling between the hole and the lattice results.

Extrapolating this picture, we expect the diminishing of the spin-Peierls distortion above the pseudogap temperature $T^*$ (note that $T^*=T_c$ for optimally doped Bi2212). As a result, the coupling between the photo-hole and the lattice is weakened. The observation of significant isotope dependences of $T^*$ (refs 27), $J$ and various low-temperature spin properties[28, 29] supports this scenario. We believe that the above cooperative interplay between electron pairing and electron–lattice interaction outlines the role that phonons play in high-temperature superconductivity. The above proposal has been shown theoretically to work in one dimension[30] and in a two-leg Hubbard ladder incorporating the electron–phonon interaction (A. Seidel, H. H. Lin and D.-H.L., manuscript in preparation).

**Acknowledgements** We are grateful to K. A. Müller, A. Bianconi, N. L. Saini, D. Pines, A. Bill, V. Z. Kresin, S. A. Kivelson, A. J. Leggett, J. Clarke, J. Orenstein, M. L. Cohen, L. Pietronero, E. Cappelluti, J. C. Davis, J. W. Allen, A. S. Alexandrov, J. C. Phillips, S. Uchida, C. Castellani, A. Bussman Holder, D. Mihailovic and C. Bernhard for discussions. We thank Z. X. Shen, Z. Hussain, D. S. Chemla and N. V. Smith for support in the initial stage of the project. The work at UC Berkeley and LBNL was supported by the Department of Energy's Office of Basic Energy Science, Division of Materials Science.

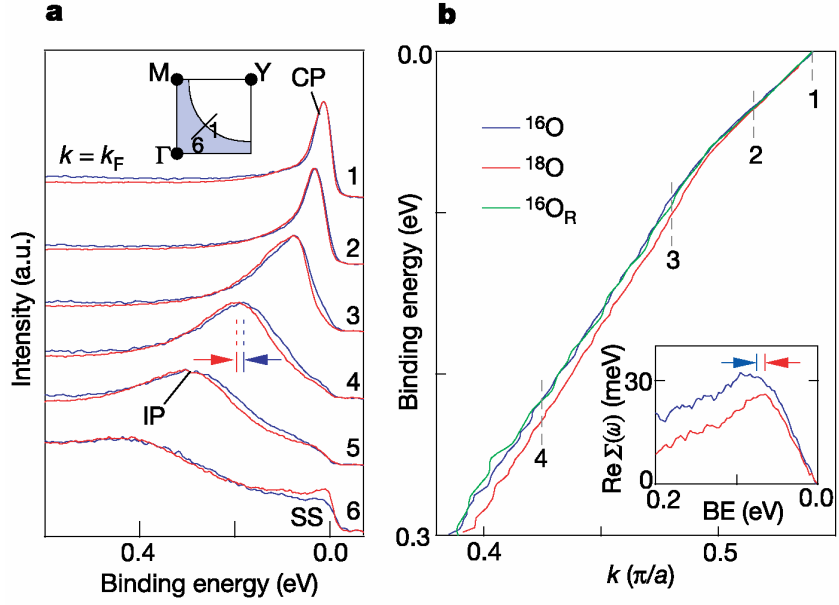

**Figure 1** Isotope-induced changes of the nodal dispersion. The ARPES data were taken on optimally doped $Bi_2Sr_2CaCu_2O_{8+\delta}$ samples with different oxygen isotopes at $T=25$ K, that is, in the superconducting state, along the nodal ($\Gamma$–Y) direction. **a**, Isotope dependence of raw EDCs. Inset shows a quadrant of the Brillouin zone, divided by the Fermi surface (curve) into electron-filled (shaded) and electron-empty parts. The momentum values corresponding to EDCs are marked in the inset and in **b** (dotted lines). The spectra show a coherent peak (CP) at low energy, almost isotope-independent, and a broad hump, that is, an incoherent peak (IP), strongly isotope dependent. The two mix in the crossover region (curve 3). Throughout this Letter, we use 'energy' and 'binding energy' interchangeably. All curves were scaled to the same peak height, and vertically displaced by different amounts for easy viewing. A small peak at the Fermi energy (0) in curve 6 is the well-known superstructure (SS) replica of the main band (also, see Supplementary Figure). **b**, Isotope dependence of MDC dispersion, obtained by lorentzian fit of MDCs. The symbol $k$ denotes the momentum value parallel to the $\Gamma$–Y direction. Consistent with **a**, the low energy dispersion is nearly isotope-independent, while the high energy dispersion is isotope-dependent. The effect is fully reversed by isotope re-substitution (green). The fits are shown only up to 300 meV, because they become less accurate at high energies. Inset shows the real part of the electron self energy, Re$S(w)$, obtained from the MDC dispersion[14] by subtracting a line approximation for one electron band $e(k)$, connecting two points, one at $E_F$ and the other at 300 meV binding energy, of the $^{18}$O dispersion. The kink position, defined as the binding energy of the peak in Re$S(w)$, undergoes a redshift upon isotope substitution (see arrows).

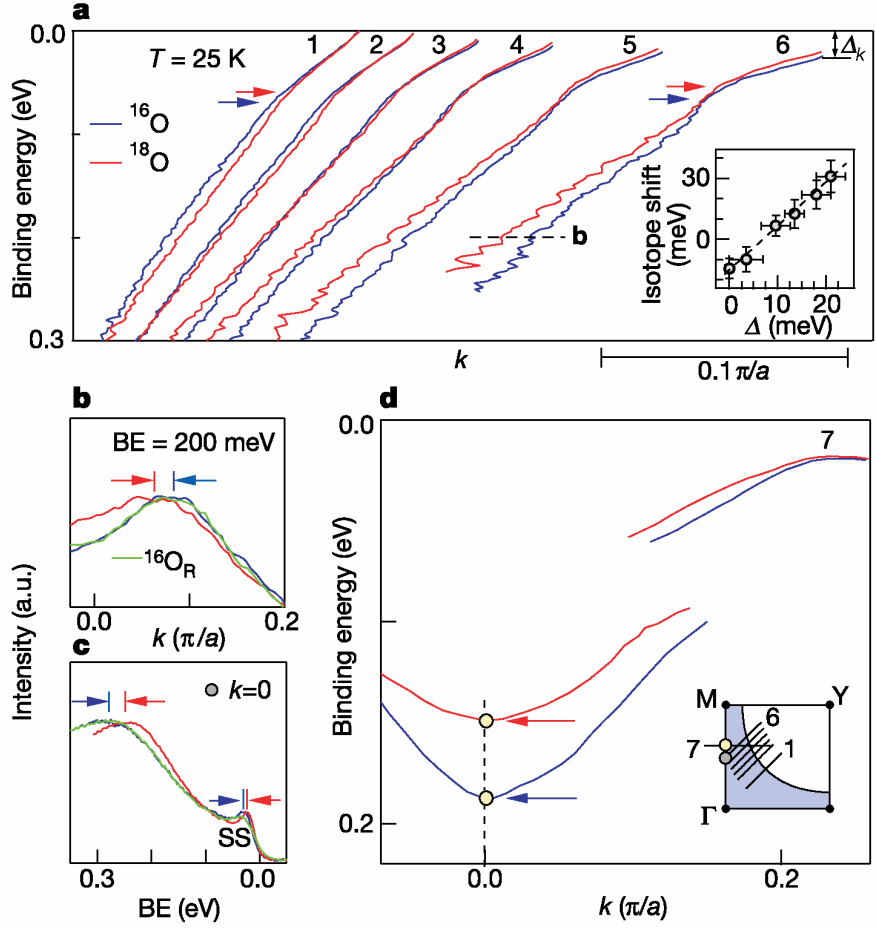

**Figure 2** Isotope-induced changes of the off-nodal dispersions in the superconducting state. The ARPES data were taken on optimally doped $Bi_2Sr_2CaCu_2O_{8+d}$ samples with different oxygen isotopes at $T=25$ K. In this figure, $k$ denotes the momentum value parallel to the M–Y axis (inset in **d**). **a**, MDC dispersions for various cuts parallel to Γ–Y (see inset in **d**). A different origin of the momentum axis is used for each cut for an easy view of all data. Inset shows the isotope shift, measured at the momentum value for which the isotope-averaged binding energy is 220 meV, versus the superconducting gap (**$D$**), isotope-averaged. The strong linear correlation (dashed line is a guide to the eye) is independent of the binding energy used. **b**, Raw MDCs at binding energy (BE) = 200 meV for the data of cut 6, confirming the fit results of **a**.. **c**, Raw EDCs at $k=0$ (see inset in **d**) for cut 6. A large reversible isotope-induced shift (about - 30 meV) of the EDC peak position is observed at high energy. Additionally, a small CP due to the well-known superstructure (SS) replica is observed (also, see Supplementary Figure). The large MDC and EDC shifts are fully reversible upon the isotope re-substitution process (green in panels **b** and **c**). **d**, EDC dispersions for cut 7 (inset), extracted from the maximum intensity positions of EDCs. The data show a large isotope-induced shift at high energy,

similarly to the data in **a**. Note also that the EDC dispersion shows a qualitatively different behaviour from the MDC dispersion, for example, those shown in **a**, namely completely separate low-energy and high-energy branches. This difference stems from the fact that near the kink EDC line shape shows a two-peak structure (for example, Fig. 2 of ref. 6) while MDC line shape continues to show a single peak.

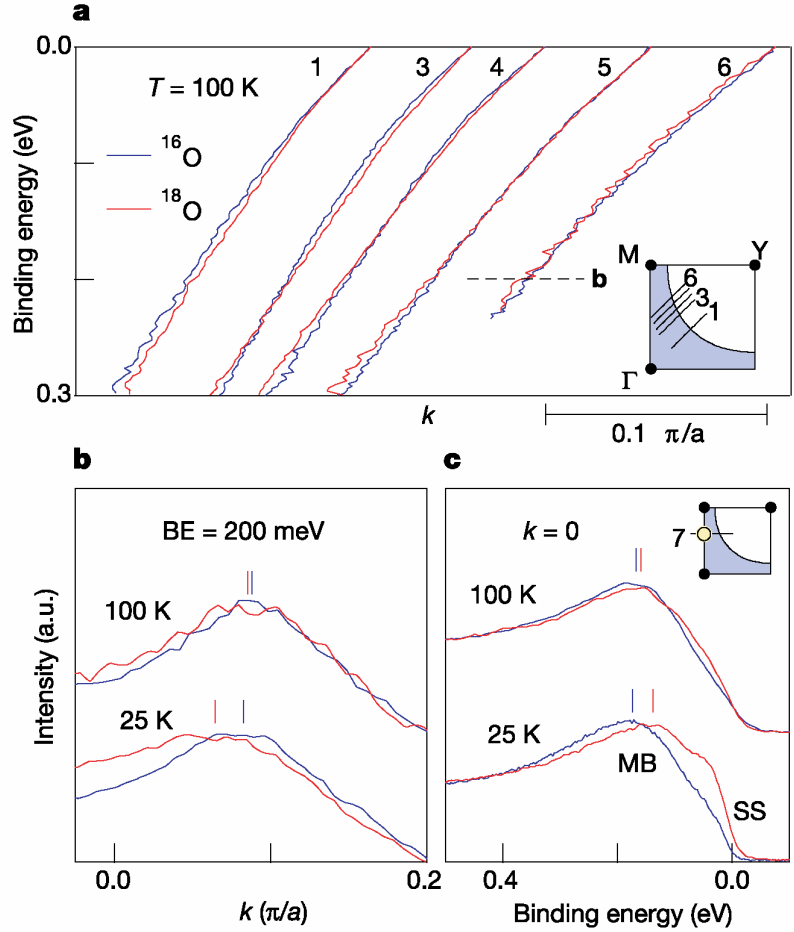

**Figure 3** Decrease of the isotope-induced changes in the normal state. The samples, the *k*-axis definition, and the location of the cuts are identical to those in Fig. 2. **a**, Normal state MDC dispersions for cuts parallel to the Γ–Y direction, from Γ–Y to near the M point, measured at *T*=100 K. A different origin of the momentum axis is used for each cut for easy viewing of all data. **b**, Comparison of the isotope-induced changes of the raw MDCs measured at 100 K and 25 K. The MDCs are measured at 200 meV binding energy for cut 6, as shown by the horizontal dashed line in **a**. **c**, Comparison of the isotope-induced changes of the raw EDCs, at 100 K and 25 K. The momentum value for the EDCs corresponds to the yellow-filled circle in the inset. As in Fig. 2c, a small peak near the Fermi energy (0) due to the well-known superstructure (SS) replica (also, see Supplementary Figure) coexists with the larger peak at high energy due to the main band (MB). An apparent isotope-induced change of the height of the SS peak at 25 K is merely due to the change of the position of the MB peak. All panels show that the isotope-induced changes are reduced greatly at high temperature, except for near the nodal direction.

**Supplementary Methods:**

**Isotope Substitution**

The oxygen isotope substitution was achieved by annealing procedures. Single crystals of $Bi_2Sr_2CaCu_2O_{8+\delta}$ were sealed into quartz tubes with an atmospheric pressure of 0.2 kbar of either $^{16}O_2$ or $^{18}O_2$ gas. They were heat-treated at the same time to insure the identical annealing condition except for the gas environment. Annealing was done at 800°C for 120 hours, followed by rapid quenching to room temperature. With refilling a fresh gas, the heat-treatment was repeated several times to increase the exchange rate. Optical reflectivity measurements on the resulting $^{16}O$, $^{18}O$, and re-substituted ($^{18}O \rightarrow {}^{16}O$) samples gave the identical spectra (plasma frequency and reflectance), indicating that the carrier concentration is the same for all three kinds of samples. The isotope substitution resulted in Raman frequency changes of only oxygen vibrations ($\omega > 250$ cm$^{-1}$), from which the exchange rate for the final $^{18}O$ sample was estimated to be more than 75% [1,2].

**Supplementary Figure:**

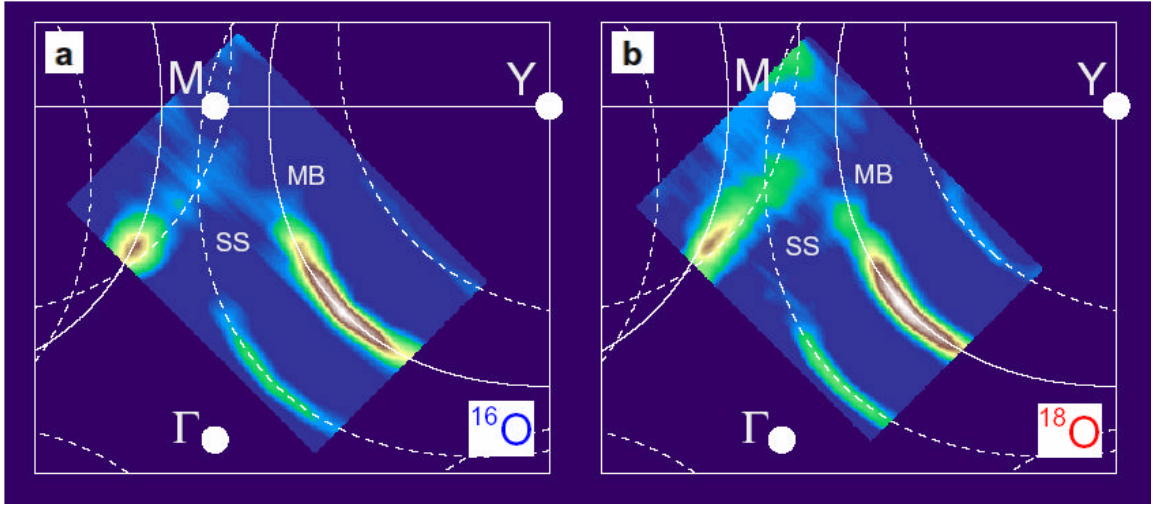

**Fig. 1 Fermi Surfaces for the two isotope samples.** Momentum space maps of the ARPES intensity integrated over an energy window of 20 meV around the Fermi energy, in the superconducting state (T = 25 K). White corresponds to the maximum intensity and dark blue to zero intensity. The maps are shown for the optimally doped $Bi_2Sr_2CaCu_2O_{8+\delta}$ samples containing oxygen isotopes $^{16}O$ (panel a) and $^{18}O$ (panel b). High intensity in these maps represents the momentum space location where low energy electronic excitations are found. As known [1,2], the spectral intensity in this energy range primarily concentrates around the node (which lies on the Γ–Y line) of the d-wave superconducting gap function. The tight-binding fits [3] of the normal state Fermi surface for the main band (MB: solid curves) and superstructure (SS) replicas (dashed curves) are shown. The difference between the tight binding parameter sets for the two samples is very small (= 3 %), indicating that the carrier concentration is unchanged by the isotope substitution. This is further supported by optical reflectivity measurements that give identical spectra (plasma frequency and reflectance [Supplementary Methods]).